\documentclass[10pt, conference]{IEEEtran}

\newcommand{\paperTitle}{Test Amplification for REST APIs \\ Using ``Out-of-the-box'' Large Language Models}

\newcommand{\submitTo}{IEEE Software -- Special Issue on Next-generation Software Testing}
\newcommand{\maxPagesAllowed}{5}

\usepackage{hyperref} 
\usepackage{amssymb} 
\usepackage{xspace}
\usepackage{color}
\usepackage{xcolor}
\usepackage{fancybox}
\usepackage{fancyhdr}
\usepackage{ifthen} 
\usepackage{paralist} 

\usepackage[T1]{fontenc}

\usepackage{comment}
\usepackage{array}
\usepackage{float}
\usepackage{adjustbox}
\usepackage{graphicx}
\usepackage{listings}
\usepackage{xcolor}
\usepackage{hyperref}
\usepackage{multirow}
\usepackage{hyperref}

\usepackage[normalem]{ulem} 

\newboolean{acmtemplate}
\setboolean{acmtemplate}{false} 
\newboolean{anonymous}
\setboolean{anonymous}{false}
\newboolean{includecomments}
\setboolean{includecomments}{true}

\newcommand{\hypobox}[1]{
	\begin{center}%
        \noindent\thicklines\setlength{\fboxsep}{2pt}%
        \cornersize{0.1}
        \ovalbox{\begin{minipage}{8.4cm}%
		#1
		\end{minipage}}
	\end{center}}

\newcommand{\tabref}[1]{Table~\ref{#1}\xspace}

\ifthenelse{\boolean{includecomments}} {
    \newcommand{\nb}[3]{
    	{\colorbox{#2}{\bfseries\sffamily\scriptsize\textcolor{white}{#1}}}
    	{\textcolor{#2}{\sf\small$\blacktriangleright$\textit{#3}$\blacktriangleleft$}}}
} {
    \newcommand{\nb}[3]{}
}

\newcommand*{\RQOne} {Isn't this a splendid research question here ?}
\newcommand*{\RQTwo} [1] {And isn't this one even better ?}
\newcommand*{\RQThree} [1] {This one tops it all, doesn't it ?}

{{\ttfamily \hyphenchar\the\font=`\-}

\clubpenalty = 10000
\widowpenalty = 10000
\displaywidowpenalty = 10000

\hypersetup{
    colorlinks,%
    citecolor=black,%
    filecolor=black,%
    linkcolor=black,%
    urlcolor=gray,
    linktocpage=true,
    pdfpagemode=UseOutlines,
    pdftitle={\paperTitle},    
    pdfauthor={Anonymous},     
    pdfsubject={Draft for \submitTo},   
    pdfcreator={PDF LaTex},   
}

\setlength{\textfloatsep}{2mm} 

\definecolor{keywordcolor}{rgb}{0.0, 0.2, 0.6} 
\definecolor{commentcolor}{rgb}{0.3, 0.6, 0.3} 
\definecolor{stringcolor}{rgb}{0.2, 0.5, 0.2} 
\definecolor{classcolor}{rgb}{0.7, 0.5, 0.0} 
\definecolor{annotationcolor}{rgb}{0.6, 0.3, 0.0} 
\definecolor{numbercolor}{rgb}{0.4, 0.4, 0.4} 

\lstdefinestyle{javacode}{
  language=Java,
  basicstyle=\ttfamily\footnotesize\color{black}, 
  keywordstyle=\color{keywordcolor}, 
  commentstyle=\color{brown}\itshape, 
  stringstyle=\color{stringcolor}, 
  classoffset=1, 
  morekeywords={UploadImage}, 
  keywordstyle=[1]\color{classcolor}, 
  keywordstyle=[2]\color{annotationcolor}, 
  morekeywords=[2]{@Test}, 
  numbers=left,
  numberstyle=\tiny\color{numbercolor}, 
  stepnumber=1,
  frame=single,
  tabsize=4,
  showstringspaces=false,
  breaklines=true,
  captionpos=b,
  numbersep=10pt, 
  framexleftmargin=15pt, 
  xleftmargin=30pt, 
}

\lstdefinelanguage{json}{
    basicstyle=\ttfamily\footnotesize\color{black}, 
    numbers=left, 
    numberstyle=\tiny\color{numbercolor}, 
    stepnumber=1, 
    numbersep=10pt, 
    showstringspaces=false, 
    breaklines=true, 
    frame=single, 
    tabsize=4, 
    framexleftmargin=15pt, 
    xleftmargin=30pt, 
    captionpos=b, 
    literate=
     *{0}{{{\color{blue}0}}}{1}%
      {1}{{{\color{blue}1}}}{1}%
      {2}{{{\color{blue}2}}}{1}%
      {3}{{{\color{blue}3}}}{1}%
      {4}{{{\color{blue}4}}}{1}%
      {5}{{{\color{blue}5}}}{1}%
      {6}{{{\color{blue}6}}}{1}%
      {7}{{{\color{blue}7}}}{1}%
      {8}{{{\color{blue}8}}}{1}%
      {9}{{{\color{blue}9}}}{1}%
      {:}{{{\bfseries\color{black}:}}}{1}%
      {,}{{{\bfseries\color{black},}}}{1}%
      {\{}{{{\color{black}\{}}}{1}%
      {\}}{{{\color{black}\}}}}{1}%
      {[}{{{\color{black}[}}}{1}%
      {]}{{{\color{black}]}}}{1}%
}

\pagestyle{fancy}
\fancyhead[CO]{REVISED -- SUBMITTED TO \submitTo}
\fancyfoot[c]{\thepage~--~\maxPagesAllowed}

\begin{document}

\pagenumbering{roman}

\newpage
\pagenumbering{arabic}
\setcounter{page}{1}

\title{\paperTitle} 

\author{
	\IEEEauthorblockN{Tolgahan Bardakci\IEEEauthorrefmark{1},
		Serge Demeyer\IEEEauthorrefmark{2},
		Mutlu Beyaz{\i}t\IEEEauthorrefmark{2}}
	\IEEEauthorblockA{
		\IEEEauthorrefmark{1}Universiteit Antwerpen
		\IEEEauthorrefmark{2}Universiteit Antwerpen and Flanders Make}
}

\IEEEtitleabstractindextext{
\begin{abstract}
REST APIs (Representational State Transfer Application Programming Interfaces) are an indispensable building block in today’s cloud-native applications, so testing them is critically important.
However, writing automated tests for such REST APIs is challenging because one needs strong and readable tests that exercise the boundary values of the protocol embedded in the REST API.
In this paper, we report our experience with using ``out of the box'' large language models (ChatGPT and GitHub's Copilot) to amplify REST API test suites.
We compare the resulting tests based on coverage and understandability, and we derive a series of guidelines and lessons learned concerning the prompts that result in the strongest test suite.
\end{abstract}

\begin{IEEEkeywords}
Rest APIs; Software Testing; Test Amplification; Artificial Intelligence; Large Language Models; Prompt Engineering
\end{IEEEkeywords}
}

\maketitle

\IEEEdisplaynontitleabstractindextext

\section{Introduction}
The API economy is an expanding trend in modern society, enabling companies and organizations to share data and functionality with other businesses, developers, and customers.
Through Application Programming Interfaces (APIs), software engineers can develop reliable applications by seamlessly integrating various components.
REST APIs, in particular, are the dominant architectural style.
Their stateless nature allows for increased scalability (e.g., auto-scaling).
Given the distributed nature of REST APIs, ensuring high quality is crucial.
Therefore, these APIs must be tested extensively.

However, testing at the API level is inherently complex.
First, there is the technical complexity induced by the sheer number of possible combinations between different protocols and an even greater number of combinations of API calls.
Secondly, different engineering teams develop different components, adding organizational complexity.
Testing boundary values remains crucial, as these will ultimately reveal the underlying defects (a.k.a. ``the needle in the haystack'').

\emph{Test amplification} is a likely solution for searching the needle in the haystack, as substantial evidence supports its effectiveness in the context of unit tests~\cite{test_amplification_ampyfier}.
Indeed, test amplifiers automatically transform an existing, manually written test suite into a more comprehensive one with stronger coverage.
An amplified test suite exercises a broader range of conditions, including boundary test values that reveal defects.

Unfortunately, the readability of the amplified tests poses a challenge.
The current generation of test amplification tools uses generic names for temporary variables (such as t1, t2, t3, \ldots).
Also, the injected code sometimes deviates from accepted coding conventions, which hinders the readability and, ultimately, the understandability of the test cases.
Using large language models for test amplification can be beneficial in addressing these issues.
Since they have seen numerous test code examples, they will likely generate meaningful names and use proper coding idioms.
However, prompt engineering is needed to optimize the test amplification process.

This paper reports using ``out-of-the-box'' large language models to amplify REST API tests.
We adopt ChatGPT 3.5~\cite{openai_chatgpt3.5}, ChatGPT 4~\cite{openai_chatgpt4}, and Copilot version 1.5.3.5510~\cite{github_copilot}.
We validate the results against a well-known representative cloud application called PetStore~\cite{petstore_application}, an open-source system with multiple API endpoints providing read, write, update, and delete actions.
We compare the results based on coverage and understandability and derive guidelines and lessons learned concerning the prompts that generate strong tests.

\section{Related Work}
\label{sec:relatedWork}
\textbf{Testing REST APIs} is crucial for ensuring the reliability and functionality of web services.
The state of the art in this domain includes a variety of techniques and tools aimed at automating and simplifying the testing process.
One key approach involves functional testing, which verifies that the API behaves as expected under various conditions.
Modern tools such as Postman, SoapUI, and RestAssured offer robust frameworks for creating and executing API tests~\cite{testing_restapis_survey}.
These automated testing frameworks have also been enhanced by continuous integration and continuous deployment (CI/CD) pipelines, allowing for more frequent and reliable testing cycles.

\textbf{To evaluate the strength} of an API test suite, a series of API coverage metrics have been proposed by Martin-Lopez et al.~\cite{test_coverage_criteria}.
These coverage metrics are defined based on elements of the OpenAPI documentation. 
They are quantified by the ratio of the number of elements observed via HTTP requests or responses to the number of elements in the documentation.

The metrics derived from observed HTTP messages include:
\begin{itemize}
\item \textbf{Path Coverage:}
The ratio of tested paths to the total documented paths.

\item \textbf{Operation Coverage:}
The ratio of tested operations to the total documented operations.

\item \textbf{Parameter Coverage:}
The ratio of input parameters to the total documented parameters.

\item \textbf{Request Content-type Coverage:}
The ratio of tested content-types to the total accepted content-types.
This excludes wildcard types (e.g., application/*).

\item \textbf{Status Code Class Coverage:}
Achieved when both correct (2XX) and erroneous (4XX, 5XX) status codes are triggered.

\item \textbf{Status Code Coverage:}
The ratio of obtained status codes to the total documented status codes.

\item \textbf{Response Content-type Coverage:}
The ratio of obtained content-types to the total response content-types documented, also excluding wildcard types.
\end{itemize}

\textbf{Test amplification} is an umbrella term for various activities that analyze and operate on existing test suites, including augmentation, optimization, enrichment, and refactoring~\cite{test_amplification_definition}.
Test amplification differs from test generation, as it creates new test cases based on existing ones instead of building them from scratch.
This provides a significant advantage as the amplified test code will better comply with the test architecture.

Artificial intelligence (AI) techniques, specifically \textbf{large language models (LLMs)}, have recently garnered a lot of attention, and numerous software engineering techniques are incorporating such AI models in various ways.
One such application is using LLMs for unit test amplification.
There are at least two recent reports concerning industrial adoption of LLMs for unit test amplification at Meta~\cite{unit_test_amplification-meta} and GitHub~\cite{unit_test_amplification_Github}.

Several AI-driven methods focus on web API testing.
RESTGPT~\cite{LLMRestApiTesting} leverages large LLMs to generate realistic, context-aware test inputs.
It directly aids API testing. 
Meanwhile, ArteLLM~\cite{ArteLLM} uses LLMs to clarify and improve API specifications.
It enhances specification-oriented testing.
However, to our knowledge, combining test amplification and LLMs to strengthen REST API tests has not yet been reported.

\hypobox{
Recently, AI tools to generate strong and readable test cases have been gaining attention.
Combining \emph{test amplification} with \emph{large language models} is particularly appealing because amplified test code will better comply with the test architecture.
However, amplifying REST API tests remains uncharted territory.
}

\section{Comparison Set-up}
Since amplifying REST API tests remains an unexplored area, we set out to investigate large language models strengthening a REST API test suite.
We restrict ourselves to ``out-of-the box'' models to obtain a minimum viable baseline.
Future extensions in particular pipelines with "Retrieval-Augmented Generation" (RAG)~\cite{lewis2021retrievalaugmentedgenerationknowledgeintensivenlp}, could go further.
We compare ChatGPT 3.5, ChatGPT 4, and Copilot version 1.5.3.5510 with varying prompts to derive guidelines and lessons learned.
Our comparison is driven by an overarching research question.
\renewcommand*{\RQOne} {How can we use large language models to amplify test code for REST APIs?}
\begin{itemize}
\item \textit{\RQOne}
\end{itemize}

We validate the results against a well-known cloud application called PetStore~\cite{petstore_application}.
Petstore is an open-source web application serving as a tutorial for deploying web services.
It provides 20 API endpoints with read, write, update, and delete operations; hence, the system is a good vehicle for experimentation.

To evaluate the quality of the amplified test code, we combine quantitative and qualitative criteria, such as structural API coverage, readability, and the amount of post-processing required. 
The detailed evaluation criteria are listed below.
All results are available in our reproduction package\footnote{\url{https://figshare.com/projects/Test_Amplification/217609}}.

\subsection{Descriptive Statistics}
To put the results in context, we count the absolute number of amplified tests.
Specifically, we tally the number of generated, successful, failed, and not applicable tests and exposed bugs.

\subsection{Structural API Coverage}
\label{sec:APIcoverage}
We use the tool Restats, written by Corradini et al.~\cite{restats} for collecting the API coverage.
The tool allows us to collect coverage metrics of every executed test case, which we combine afterward in tabular form.
Specifically, we use the following metrics, a subset of the ones defined by Martin-Lopez et al.~\cite{test_coverage_criteria}.
\begin{itemize}
\item{Path Coverage}
\item{Operation Coverage}
\item{Status Class Coverage}
\item{Status Coverage}
\item{Response Type Coverage}
\item{Request Type Coverage}
\item{Parameter Coverage}
\end{itemize}

\subsection{Amount of Post-Processing}
\label{sec:postprocesing}
In git-based software engineering environments, code changes are submitted via pull requests.
This implies that some human post-processing is needed before the changes are accepted into the code base.
We mimic this by manually reviewing the amplified test code and making slight alterations when needed.
The purpose of these alterations is to bring the test suite into an executable form.
As a proxy measure for the amount of work this entails, we count the number of lines edited.

\subsection{Readability}
\label{sec:readability}
Besides the above quantitative evidence, we adopt one qualitative criterion.
We assess the readability of the amplified test code using the following questions.
The questions are answered by the first author and reviewed by the two other authors.
\begin{itemize}
\item{Are the tests understandable from the human perspective?}
\item{Do they include appropriate comments?}
\item{Do they comply with the common coding idioms?}
\item{Are they really useful in a way that few or no edits are required for clarity?}
\end{itemize}

\section{Comparison}
We start from a happy-path test script for one endpoint (\texttt{/pet/\{petId\}/uploadImage}), which uploads an image to a given pet ID.

The test script is shown in \autoref{lst:uploadImage}.

\begin{lstlisting}[style=javacode, caption={Happy-Path Test Script}, label={lst:uploadImage}]
@Test
public void uploadImageHappyPath () {
	String formData = "../../data/tolgahanimage";
	int petId = 2;
	Response response = post("/pet" + "/" + petId +"/uploadImage", null, null, formData, null, null);
	Assert.assertEquals(response.getStatusCode(), 200);
	}
\end{lstlisting}

Starting from this baseline, the actual comparison is driven by increasingly stricter prompts executed against the three LLMs under investigation: ChatGPT 3.5, ChatGPT 4, and Copilot.
\begin{itemize}
\item \textbf{Prompt 1.}
Our first prompt is the simplest thing that could possibly work.
We provide the happy-path test script and ask the LLM ``Can you perform test amplification?''
Sometimes, the LLM does not return any test code but instead provides test scenarios in natural language. 
In those instances, we follow up with a question similar to ``Can you write the test code for these scenarios?''
\item \textbf{Prompt 2.}
For the second prompt, we provide the OpenAPI documentation as an extra input, expecting tests with better coverage.
\item \textbf{Prompt 3.}
With the third prompt, we ask the LLMs for the maximum number of test cases they can amplify, in addition to the happy-path test scenario and OpenAPI documentation.
We place it at the end of the prompt after the happy-path test and OpenAPI documentation.
\end{itemize}

\subsection{Descriptive Statistics}

\tabref{tab:prompt_statistics} shows how many tests were created by the different prompts.
As expected, Prompt 3 (maximize the amount of tests) generates the most tests.
Prompt 2 (add the API documentation) has the most impact on Copilot: 12 tests instead of 3.

\begin{table}[H]
\centering
\caption{Descriptive Statistics for Different Prompts}
\label{tab:prompt_statistics}
\resizebox{0.50\textwidth}{!}{
\begin{tabular}{|c|l|c|c|c|}
\hline
\multicolumn{1}{|c|}{} & \textbf{Statistic} & \textbf{GPT 3.5} & \textbf{GPT 4} & \textbf{Copilot} \\ \hline
\multicolumn{5}{|c|}{} \\ \hline
\multirow{7}{*}{\textbf{Prompt 1}} 
& Generated Tests             			& 7   	& 8   	& 3   \\ \cline{2-5}
& Successful Tests         				& 5   	& 4   	& 2   \\ \cline{2-5}
& Failed Tests     					& 2   	& 4   	& 1   \\ \cline{2-5}
& N/A (Not Acceptable Tests) 			& 0   	& 0   	& 0   \\ \cline{2-5}
& Bugs Exposed 					& 1   	& 4   	& 1   \\ \cline{2-5}
& Post-Processing (No. of lines edited)	& 3   	& 1   	& 0   \\ \hline
\multicolumn{5}{|c|}{} \\ \hline
\multirow{7}{*}{\textbf{Prompt 2}}
& Generated Tests              			& 5   	& 8   	& 12		\\ \cline{2-5}
& Successful Tests         				& 5   	& 4   	& 10   	\\ \cline{2-5}
& Failed Tests     					& 0   	& 2   	& 2   		\\ \cline{2-5}
& N/A (Not Acceptable Tests)            		& 0   	& 2   	& 0	  	\\ \cline{2-5}
& Bugs Exposed					& 0   	& 2   	& 2   		\\ \cline{2-5}
& Post-Processing (No. of lines edited)	& 0   	& 7   	& 13   	\\ \hline
\multicolumn{5}{|c|}{} \\ \hline
\multirow{7}{*}{\textbf{Prompt 3}}
& Generated Tests              			& 15   	& 9		& 17		\\ \cline{2-5}
& Successful Tests         				& 14		& 7   		& 16   	\\ \cline{2-5}
& Failed Tests      					& 0		& 2  		& 0	   	\\ \cline{2-5}
& N/A (Not Acceptable Tests)            		& 1		& 0   		& 1   		\\ \cline{2-5}
& Bugs Exposed     					& 0		& 2   		& 0   		\\ \cline{2-5}
& Post-Processing (No. of lines edited)	& 52		& 22		& 30   	\\ \hline
\end{tabular}
}
\vspace{0.5em}
\end{table}

On rare occasions, we encounter "Not Acceptable" test cases.
These cases exercise "Deprecated Endpoints".
Therefore, they are neither "successful" nor "failed".
LLMs created these test cases mainly because they were documented in the OpenAPI specification.
However, in practice, they have become deprecated in the cloud.

The row ``Bugs Exposed'' is quite insightful as well.
To illustrate what is happening, we use an example test created by ChatGPT 4 in \autoref{lst:gpt4amplified_test}.
This test exercises the end-point with incorrect input, forcing the API under test to be in a special erroneous state and expecting error status codes.
When we run the amplified test (\autoref{lst:gpt4amplified_test}), we deploy an invalid \texttt{petId}, and we do not expect a 200 status code (line 7).
However, the JSON output is shown in \autoref{lst:bug_json}, and lines 2 and 7 show it is a status code 200.
This is an example of a test case that exposes a bug in the API under test.

GPT 4.0 had the most bug-exposing tests (four in Prompt 1 and two in Prompts 2 and 3).
However, GPT 3.5 and Copilot had a few of these as well.

\begin{lstlisting}[style=javacode, caption={Amplified Test Script by GPT 4}, label={lst:gpt4amplified_test}]
// 3. Test with Invalid Pet ID
@Test
public void uploadImageInvalidPetId() {
	String formData = "../../data/tolgahanImage";
	int petId = -1; // Assuming -1 is an invalid ID
	Response response = post("/pet" + "/" + petId + "/uploadImage", null, null, formData, null, null);
	Assert.assertNotEquals(response.getStatusCode(), 200);
	}
\end{lstlisting}

\begin{lstlisting}[language=json, caption={Bug Example JSON}, label={lst:bug_json}]
{
	"code": 200,
	"type": "unknown",
	"message": "additionalMetadata: null\nFile uploaded to ./null, 24 bytes"
}

java.lang.AssertionError: did not expect [200] but found [200]
\end{lstlisting}

\subsection{Structural API coverage}
\tabref{tab:prompt_coverage} shows the respective impact on the coverage metrics for the successive prompts and compares with the baseline.
Prompt 1 has no impact on the various coverage metrics, \textsf{Status} and \textsf{Status Class} being the noteworthy exceptions.
The simplest prompt generates tests that make the REST API return correct (2XX) and erroneous (4XX) status codes.
GPT 4.0 was smarter than the other two in the sense that it did not increase the coverage but instead exposed bugs, as discussed previously.

The API specification provided in Prompt 2 had a significantly positive impact.
Copilot increases the coverage significantly for all coverage metrics.
GPT 4.0 (and to a lesser extent GPT 3.5) sees only changes for the \textsf{Status}, \textsf{Status Class}, and \textsf{Parameter}.
But this illustrates quite well how the API specification permitted LLMs to create stronger tests.

Prompt 3 delivers the best results regarding coverage. 
A notable observation is the significant increase in \textsf{Path Coverage} across all three LLMs. 
This suggests that additional tests exercise different endpoints.
Thus, combining an example test script (Prompt 1) with the API specification (Prompt 2) and requesting to maximize the number of test cases (Prompt 3) creates tests for the whole API.

\begin{table}[H]
\caption{API Coverage for Different Prompts}
\label{tab:prompt_coverage}
\begin{tabular}{|c|l|c|c|c|}
\hline
\multicolumn{1}{|c|}{} & ~ & \textbf{Coverage} & ~ & ~ \\ \hline
\multirow{7}{*}{\textbf{Baseline}} 
& Path              		& 7\%   	& ~   	& ~   \\ \cline{2-5}
& Operation         	& 5\%   	& ~   	& ~   \\ \cline{2-5}
& Status Class     	& 4\%   	& ~   	& ~   \\ \cline{2-5}
& Status 			& 3\%   	& ~   	& ~   \\ \cline{2-5}
& Response Type 	& 3\%   	& ~   	& ~   \\ \cline{2-5}
& Request Type	& 9\%   	& ~   	& ~   \\ \cline{2-5}
& Parameter         	& 11\%  	& ~  	& ~  \\ \hline
\multicolumn{1}{|c|}{} & ~& \textbf{GPT 3.5} & \textbf{GPT 4} & \textbf{Copilot} \\ \hline
\multirow{7}{*}{\textbf{Prompt 1}} 
& Path              		& 7\%   	& 7\%   	& 7\%   \\ \cline{2-5}
& Operation         	& 5\%   	& 5\%   	& 5\%   \\ \cline{2-5}
& Status Class     	& 7\%   	& 4\%   	& 7\%   \\ \cline{2-5}
& Status 			& 8\%   	& 3\%   	& 5\%   \\ \cline{2-5}
& Response Type 	& 3\%   	& 3\%   	& 3\%   \\ \cline{2-5}
& Request Type	& 9\%   	& 9\%   	& 9\%   \\ \cline{2-5}
& Parameter         	& 11\%  	& 11\%  	& 11\%  \\ \hline
\multicolumn{5}{|c|}{} \\ \hline
\multirow{7}{*}{\textbf{Prompt 2}}
& Path              		& 7\%   	& 7\%   	& 64\%   \\ \cline{2-5}
& Operation         	& 5\%   	& 5\%   	& 55\%   \\ \cline{2-5}
& Status Class     	& 4\%   	& 7\%   	& 18\%   \\ \cline{2-5}
& Status            	& 3\%   	& 5\%   	& 13\%   \\ \cline{2-5}
& Response Type	& 3\%   	& 3\%   	& 29\%   \\ \cline{2-5}
& Request Type	& 9\%   	& 9\%   	& 45\%   \\ \cline{2-5}
& Parameter		& 11\%  	& 22\%  	& 33\%  \\ \hline
\multicolumn{5}{|c|}{} \\ \hline
\multirow{7}{*}{\textbf{Prompt 3}}
& Path              		& 71\%   	& 43\%	& 93\%   \\ \cline{2-5}
& Operation         	& 80\%  	& 35\%   	& 85\%   \\ \cline{2-5}
& Status Class      	& 26\%	& 11\%  	& 26\%   \\ \cline{2-5}
& Status            	& 19\%	& 8\%   	& 19\%   \\ \cline{2-5}
& Response Type     	& 42\%	& 18\%   	& 45\%   \\ \cline{2-5}
& Request Type      	& 64\%	& 45\%   	& 73\%   \\ \cline{2-5}
& Parameter         	& 33\%	& 33\%  	& 78\%  \\ \hline                                         
\end{tabular}
\end{table}

\subsection{Post-processing}
Looking at the rows \textsf{Post-Processing} in \tabref{tab:prompt_statistics}, we observe that when only a small amount of test code is generated, minimal post-processing is necessary.
However, more editing is required with increasing tests (with Prompts 2 and 3).
The increased number of edited code lines is not simply due to the increased number of tests.
Since we only provide one test case for one endpoint, the LLMs generate relatively short test cases.
Still, they generate tests for other endpoints, especially with Prompt 3.
Consequently, they make minor errors that need human attention while producing additional test cases.

For GPT 4 and Copilot, through the more advanced prompts, it increases steadily.
However, for GPT 3.5, if we look at Prompt 3 on \tabref{tab:prompt_statistics}, it increases significantly compared to the previous prompts.
It requires 2.4 times more effort than GPT 4 and 1.7 times more effort than Copilot.

\subsection{Readability}
One of the biggest advantages of using LLMs is that they are built to generate meaningful text. 
For all the LLMs, GPT 3.5, GPT 4, and GitHub Copilot, it is true that, in the produced test code, the test method names and the variable names are excellent.
In addition to that, the code lines are clear, and the comments are understandable.
They follow coding idioms and generate meaningful method names such as "deleteOrderInvalidId".

\section{Future Work}
The following points are potential explorations for the future.

	\begin{itemize}
	\item{
	Our investigation focused specifically on REST APIs, whereas previous work focused on unit testing.
	Given the promising results, we should consider other testing contexts, such as web UI and mobile application testing.
	}
	\item{
	Our prompt strategies were designed to maximize the strength of the test suite via structural coverage metrics.
	However, other goals might be expressed (e.g., maximizing bug exposure), in which case, different prompt strategies would be appropriate.
	}
	\item{
	Validating these findings through industry partner case studies would be highly beneficial.
	This real-world aspect would uncover insights into the strengths and weaknesses of this approach.}
	\item{
	We seeded the LLM with a one happy-day test case for a single API end-point.
	In reality, a test suite for a REST API will consist of several tests, possibly end-to-end tests representing user stories.
	It remains to be seen whether LLMs can amplify a complete test suite.
	}
	\item{
	LLMs are rapidly developing; we may soon expect more specialized versions trained for writing test code.
	One possible approach would be using RAG (Retrieval Augmented Generation) models for better results.
	Replicating our work with specialized LLMs and using more advanced prompting techniques may reveal interesting insights.}
	\end{itemize}

\section{Threats to the Validity}
As with all empirical research, some factors may jeopardize the validity of our results.
Below, we list those factors and our actions to reduce or alleviate the risks. 

\begin{enumerate}
\item{
The outputs of LLMs may vary over time, even when given the same prompt.
This is a known problem with the use of LLMS.
In this case, we do not expect the variation to affect the amount of test cases generated or the Path or Status Class coverage.
For the other criteria, we only expect a minor impact.
}
\item{As we use an open-source cloud application, which has been used in other test generation experiments, the LLMs may have already seen valid test cases.
We intend to replicate the investigation with other REST APIs to verify whether this is an issue.}

\item{
The calculation of API coverage was a semi-automatic process; hence, human error is possible.
To minimize that risk, we conducted a pilot study using a simple REST API (with only read access at a single entry point.)
}
\end{enumerate}

From a practical standpoint, there are validity issues when considering real-world applications:
\begin{enumerate}
\item{Due to privacy concerns, some company policies may prohibit using LLMs in software development.
Explicitly submitting the OpenAPI documentation to an LLM indeed induces a security risk.
We warn prospective users about these potential policy violations and argue that they should check beforehand.}
\item{
There are many REST APIs, and style and complexity may vary greatly.
We have attempted to select an appropriate cloud application with all REST API operations, but other cloud applications may yield different results.
Investigating other REST APIs should minimize that risk.
}
\end{enumerate}

\section{Conclusion}
This comparison illustrates how test amplification combined with LLM could strengthen REST API test suites.
When asked to amplify an existing happy-day test, all LLMs created extra API tests.
Extra API tests have increased coverage and resulted in readable test cases that require little post-processing to be accepted as pull requests.
Some extra tests even exposed bugs, illustrating that such tests can find the proverbial "needle in a haystack."
The provided prompt significantly impacts coverage (see \tabref{tab:prompt_statistics}).
However, when we exercise only one endpoint and create amplified tests (Prompt 1), models tend to provide detailed coverage of boundary values and test cases.
Providing the appropriate additional information in the prompt ---such as the OpenAPI specification (prompt 2) or asking for the maximum number of tests (prompt 3)--- significantly improves the results for GPT 4 and Copilot.
When LLMs have more information, they not only improve the coverage but ---more importantly--- also exercise other endpoints.

\section{Acknowledgments}
This work is supported by the Research Foundation Flanders (FWO) via the \href{https://soft.vub.ac.be/basecampzero/index.html}{BaseCamp Zero Project} under Grant number S000323N.

\newpage 

\ifthenelse{\boolean{acmtemplate}} {
	\bibliographystyle{ACM-Reference-Format}
}{
	\bibliographystyle{IEEEtran}
}
\bibliography{bardakci2024ieeeBib}

\end{document}